\documentclass[sigconf]{acmart}
\AtBeginDocument{%
  }

\copyrightyear{2025}
\acmYear{2025}
\setcopyright{acmlicensed}
\acmConference[WWW Companion '25] {Companion of the 16th ACM/SPEC International Conference on Performance Engineering}{April 28-May 2, 2025}{Sydney, NSW, Australia.}
\acmBooktitle{Companion of the 16th ACM/SPEC International Conference on Performance Engineering (WWW Companion '25), April 28-May 2, 2025, Sydney, NSW, Australia}
\acmISBN{979-8-4007-1331-6/25/04}
\acmDOI{10.1145/3701716.3715575}
\begin{document}

% \title{Assessing the probability of State-State coordination on Twitter}
\title[Assessing Claims of Coordinated Inter-State Information Operations on Twitter/X]{Beyond Interaction Patterns: Assessing Claims of Coordinated Inter-State Information Operations on Twitter/X}
% \title{Beyond Interaction Patterns: Assessing Claims of Coordinated Inter-State Information Operations on Social Media}


%         Issues to address from reviewers
% 1) Greater discription of the control data used
% 2) Greater specification of biases potenitally present in control data
% 3) List results of tests fully

% Acknowledge help from Filipi

\author{Valeria Pantè}
\authornote{These authors contributed equally to this work. Author emails: Valeria Pantè, \texttt{valeriapante@gmail.com}; David Axelroad, \texttt{daaxelro@iu.edu}.}
\affiliation{
\institution{Information Sciences Institute, University of Southern California}
  \city{Los Angeles}
  \state{California}
  \country{USA}
  \orcid{0009-0000-4440-331X}
}

\author{David Axelrod}
\authornotemark[1]
\affiliation{
\institution{Observatory on Social Media, Indiana University}
  \city{Bloomington}
  \state{Indiana}
  \country{USA}
  \orcid{0000-0002-9840-8711}
}

\author{Alessandro Flammini}
\affiliation{
\institution{Observatory on Social Media, Indiana University}
  \city{Bloomington}
  \state{Indiana}
  \country{USA}
  \orcid{0000-0003-1670-9156}
}

\author{Filippo Menczer}
\affiliation{
\institution{Observatory on Social Media, Indiana University}
  \city{Bloomington}
  \state{Indiana}
  \country{USA}
\orcid{0000-0003-4384-2876}
}

\author{Emilio Ferrara}
\affiliation{%
  \institution{Information Sciences Institute, University of Southern California}
  \city{Los Angeles}
  \state{California}
  \country{USA}
  \orcid{0000-0002-1942-2831}
}

\author{Luca Luceri}
\affiliation{%
  \institution{Information Sciences Institute, University of Southern California}
  \city{Los Angeles}
  \state{California}
  \country{USA}
  \orcid{0000-0001-5267-7484}
}

\renewcommand{\shortauthors}{Valeria Pantè et al.}
%% No italics, no superscripts
%% Use footnote or author note to identify equal contribution and/or contact author info

\begin{abstract}
  % Past studies examining the Twitter Information Operation archive have highlighted interactions between actors belonging to coordination efforts from different states. This has led to speculation that these interacting states may be coordinating their activities. In this study we leverage data collected as a control for  campaigns in the IO archive and evaluate if interactions between states are statistically probable given expectations set by the control data. We evaluate the potential for coordination with respect to sharing similar content as well as directly retweeting or replying to each other. We find that no pair of countries systematically exceed control levels of similarity or interactions.
  Social media platforms have become key tools for coordinated influence operations, enabling state actors to manipulate public opinion through strategic, collective actions. While previous research has suggested collaboration between states, such research failed to leverage state-of-the-art coordination indicators or control datasets. In this study, we investigate inter-state coordination by analyzing multiple online behavioral traces and using sophisticated coordination detection models. By incorporating a control dataset to differentiate organic user activity from coordinated efforts, our findings reveal no evidence of inter-state coordination. These results challenge earlier claims and underscore the importance of robust methodologies and control datasets in accurately detecting online coordination.
\end{abstract}



%  \centering
%     \includegraphics[width=0.7\textwidth]{images/graphs_norm.png}
%     \caption{Inter-country connections based on different similarity features. Edges between countries represent the strength of their relationship, quantified by Jaccard similarity which is computed as the ratio of shared features to the total number of unique features between the two countries.}
%     \label{fig:graphs}


\maketitle

\section{Introduction}

Social media has fundamentally changed the ways in which users connect, communicate, and influence each other across borders. However, alongside these benefits, social platforms have also become vehicles for coordinated online influence operations (IO), where malicious actors engage in organized efforts to manipulate public opinion or sway perceptions through strategic, collective actions. Since the lead-up to the 2016 U.S. Presidential Election, a body of literature has documented instances of state-run campaigns on social media during geo-political events \cite{badawy2018characterizing, zannettou2019disinformation, saeed2022troll, fisher2020demonizing}, including during the 2024 U.S. election \cite{minici2024uncovering,cinus2024exposing}. 
These campaigns include the deployment of automated accounts (bots) and state-sponsored human operators (trolls) to disseminate or amplify content across multiple social media platforms, with goals ranging from election interference to managing state public relations \cite{jacobs2022who, jacobs2023what}.

To effectively propagate their chosen narratives, these campaigns often rely on coordinated actions that mimic organic virality, aiming to generate sufficient synthetic activity to create the illusion of widespread agreement and encourage organic users to adopt certain viewpoints \cite{coordination}. Coordinated actors utilize various tactics to advance their agenda, often aiming to manipulate a platform's feed recommendation algorithm or co-opting influential users to promote an agenda to their sizable audiences \cite{Pote2024replies,nwala2023language,suresh2023tracking}. 

To counter this growing threat, numerous computational models have been developed to detect coordinated actions by state-backed users. Many of these approaches focus on identifying improbable degrees of similarity in various user behaviors, suggesting that these users are not acting independently \cite{Pacheco_2020, Pacheco_2021, luceri2024unmasking, nizzoli2021coordinated, weber2021amplifying, magelinski2022synchronized}. The assumption is grounded in the premise that social media interactions typically reflect independent user behavior. Deviations from this behavioral pattern, characterized by abnormal similarity in online activities, may signal coordinated inauthentic behavior. 
Similar assumptions are implicit in rules about coordinated inauthentic behavior on platforms such as X\footnote{help.x.com/en/rules-and-policies/authenticity} and Facebook.\footnote{about.fb.com/news/tag/coordinated-inauthentic-behavior/} 
Such coordination may involve sharing identical content, synchronized re-shares, artificial amplification, or the promotion of specific keywords, hashtags, or URLs \cite{Pacheco_2020, giglietto2020takes, coURL,Pacheco_2021, burghardt2023socio, synchrotrap,suresh2023tracking,debot, magelinski2022synchronized, tardelli2023temporal}. In addition to approaches that focus on individual behaviors, efforts have been made to analyze multiple behavioral traces within a unified model that combines coordination indicators \cite{luceri2024unmasking,syncActionFrame,multiviewClustering,luceri2024leveraging,vargas2020detection,sharma2021identifying,erhardt2023hidden}. 

Recently, Wang et al.~\cite{wang2023evidence} extended the analysis beyond the user level to examine whether coordination can be observed at the state level. By aggregating users controlled by a single state entity, they constructed state-state interaction networks, reporting that states may collaborate to promote shared agendas. The authors report identifying signals of \textit{``intentional, strategic interaction amongst thirteen different states, separate and distinct from within-state operations.''} While their study provides valuable insights into potential coordination between states, the methodology focuses only on user interaction patterns and does not incorporate state-of-the-art approaches for detecting coordination.

%\subsection*{Contribution of this work}

In this study, we investigate potential inter-state coordinated influence campaigns by analyzing multiple online behavioral traces and leveraging state-of-the-art coordination detection models to identify potential strategic joint efforts among different states.
Unlike Wang et al.~\cite{wang2023evidence}, we enhance our analysis by incorporating a control dataset that captures the activity of organic users over the same time frame and discussing similar topics to those addressed by coordinated actors. By examining behavioral traces such as interactions through replies and retweets, or similarity patterns like co-retweets, co-hashtags, co-URLs, rapid retweets, and text similarity, we aim to identify whether these behaviors indicate systematic coordination beyond organic social media interactions.
To this end, we employ network-based models \cite{Pacheco_2020,luceri2024unmasking} and statistical testing to assess whether the observed inter-state interactions and similarities are unlikely to occur by chance.
Our results show no statistical evidence of coordination among different states, failing to replicate the findings of Wang et al.~\cite{wang2023evidence}. These findings underscore the limitations of approaches that rely solely on interaction patterns and do not integrate robust control datasets to detect online coordination.

% \section*{Related Work}

% To date, user behaviors analyzed for suspicious similarities include the sharing of URLs or hashtags \cite{Pacheco_2020, giglietto2020takes, coURL,Pacheco_2021, burghardt2023socio}, re-sharing the same posts or users, operating with temporal synchronicity \cite{Pacheco_2021, synchrotrap,suresh2023tracking,debot, magelinski2022synchronized, tardelli2023temporal}, and posting content with very similar text \cite{nizzoli2021coordinated, Pacheco_2020,suresh2023tracking}. Pacheco et al., \cite{Pacheco_2021} also leverage suspicious account handle sharing behavior to identify coordinated accounts. 
% In addition to examining the collective behavior of accounts, some work has also looked at individual account traces that betray some degree of account automation \cite{ferrara2016rise,badawy2018analyzing,mazza2022investigating}. Use of software to automate account behavior has some benign applications, but it also can be critical to running an information operation with limited human resources. In addition to approaches that consider one behavior of the above behaviors, attempts have been made to analyze mutliple user behaviors within a single model \cite{syncActionFrame,multiviewClustering,weber2021amplifying,vargas2020detection,sharma2021identifying,luceri2024unmasking,erhardt2023hidden}. 

\section*{Data}

To promote transparency and support research, social media platforms like Twitter\footnote{web.archive.org/web/20240829231920/https://transparency.x.com/en/reports/moderation-research} (now X) and Facebook\footnote{transparency.meta.com/metasecurity/threat-reporting} have maintained archives of IOs detected and attributed to state actors. Accounts involved in these operations were suspended for violating the platform's terms of service,\footnote{help.twitter.com/en/rules-and-policies/platform-manipulation} which define platform manipulation as ``activity that attempts to artificially influence conversations through the use of multiple accounts, fake accounts, automation, and/or scripting,'' aiming to ``make accounts or content appear more popular or active than they are.'' 
%When Twitter identified coordinated inauthentic activity originating from state actors, the accounts were suspended and their posts archived.

We utilize a combination of IO and control datasets collected for 26 influence campaigns originating from 16 different countries \cite{seckin2024labeled}. The control data captures organic user activity during the same time frame as the coordinated IO actors and focuses on the same topics. Table~\ref{table:users_count} summarizes the number of users involved for each country and class of account (IO vs. control).

 %\vspace{0.5cm}
  \begin{table}
  %\begin{minipage}[t!]{\columnwidth}
  %\vspace{0pt}
    \centering
    \begin{tabular}{lcc}
    \hline
    \textbf{State} & \textbf{IO users} & \textbf{Control users}\\ \hline   
    Armenia & 31 & 1,767\\ %\cline{1-3}
    Iran & 5,902 & 37,232\\ %\cline{1-3}
    Ghana \& Nigeria & 60 & 1,166\\ %\cline{1-3}
    Ecuador & 787 & 21,138\\ %\cline{1-3}
    Catalonia & 76 & 2,607\\ %\cline{1-3}
    China & 890 & 76,286\\ %\cline{1-3}
    Cuba & 503 & 30,099\\ %\cline{1-3}
    Thailand & 455 & 2,549\\ %\cline{1-3}
    Russia & 3,918 & 80,108\\ %\cline{1-3}
    Venezuela & 611 & 9,745\\ %\cline{1-3}
    Qatar & 29 & 19,481\\ %\cline{1-3}
    Egypt \& UAE & 3,898 & 10,876\\ %\cline{1-3}
    Bangladesh & 11 & 929\\ %\cline{1-3}
    Spain & 216 & 1,681\\ %\cline{1-3}
    \hline
    \end{tabular}
    %\captionof{table}{Number of users involved in each country}
    \caption{Number of users involved in each country.}
    \label{table:users_count}
    %\end{minipage}
    \end{table}
%\vspace{-0.5cm}

\begin{figure*}[t!]
    \centering
    \includegraphics[width=0.99\textwidth]{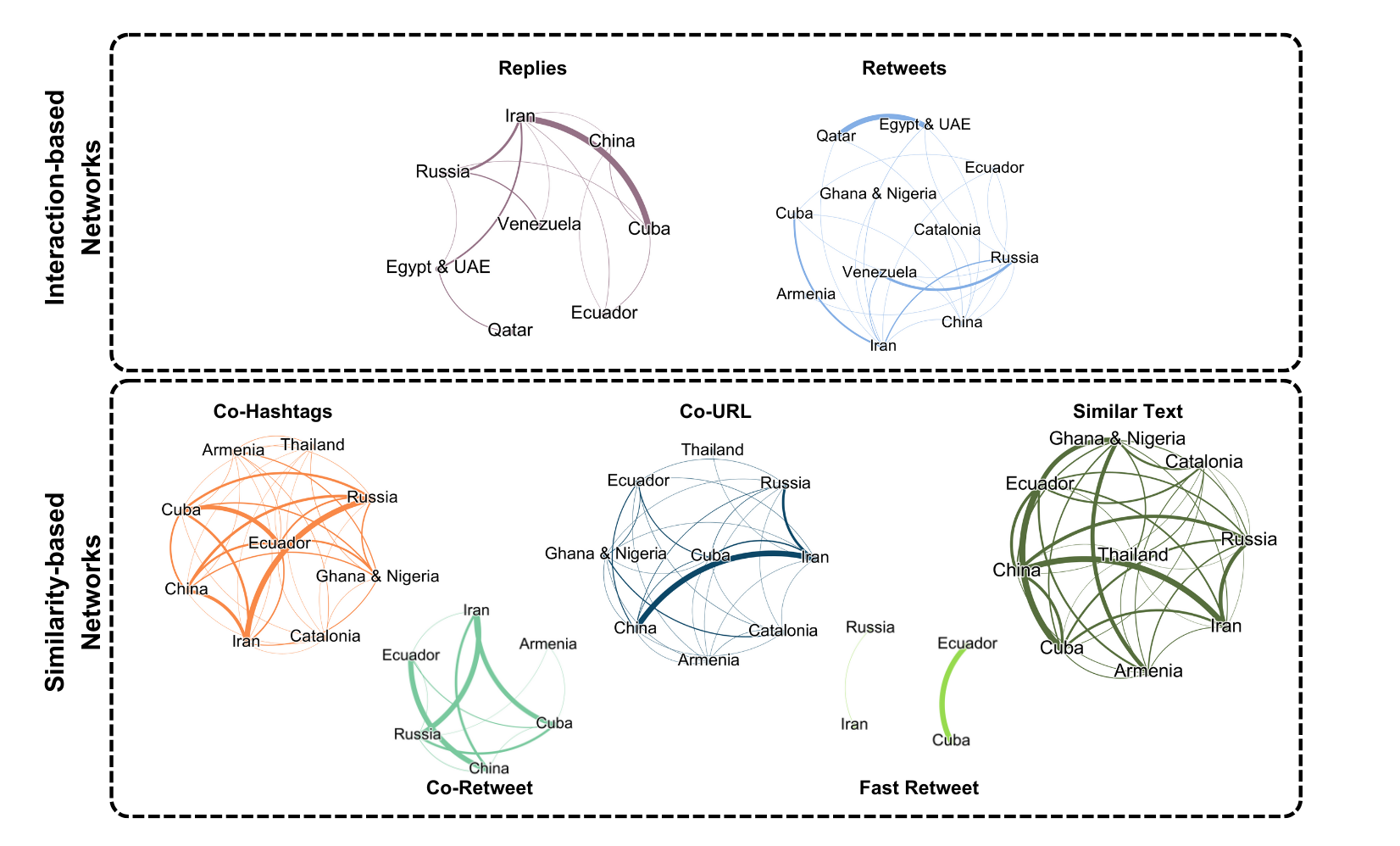}
    \caption{Inter-country connections based on aggregated interactions (top row) and similarity features (bottom row) among IO users. For interactions, edges are weighted by the out-strength of each node. For similarity networks, edges are weighted by the Jaccard coefficient, which is computed as the ratio of shared features to the total number of unique features between the two countries.}
    \label{fig:graphs}
\end{figure*}

\section*{Methods}

\subsection*{Similarity-Based Coordination Detection}

The traditional approach to detecting online coordination on social media relies on the assumption that authentic users act independently, exhibiting limited similarities in their online activities \cite{Pacheco_2020}. An anomalously high similarity among users may signal \textit{similarity-based coordination} \cite{Pacheco_2021}. 
%Similarly to \cite{luceri2024unmasking}, 
To identify such coordination, we analyze five distinct behavioral traces: \textit{Co-Retweet}, i.e., re-sharing the same tweets; \textit{Co-URL}, i.e., sharing the same URLs;
\textit{Co-Hashtag}, i.e., using the same hashtags;
\textit{Fast Retweet}, i.e., quickly re-sharing content from the same source account within a short time frame; and \textit{Text Similarity}, i.e., posting content with similar textual content.

We represent users 
%as normalized vectors 
based on their activity patterns. 
Specifically, for each similarity indicator, users are represented by term frequency-inverse document frequency (TF-IDF) 
%normalized count 
vectors. 
The vector dimensions are tweet IDs, URL domains, and hashtags for the co-retweet, co-URL, and co-hashtag networks, respectively. 
We then construct a user similarity network 
%that capture the pairwise similarity of users engaging in 
for each behavior by weighting the edge between each pair of users according to the cosine similarity between their vectors \cite{Pacheco_2021,Pacheco_2020,luceri2024unmasking}. 

A similar procedure is applied for the fast-retweet and text-similarity user networks, albeit with some preprocessing steps. The construction for fast-retweets is as outlined above for co-retweet, but we include only retweets that occur within the first 10 seconds after a tweet is posted.
%, then the construction of the similarity network follows the mechanism . 
For text-similarity, we only consider original posts and remove punctuation, stopwords, emojis, URLs, hashtags, and mentions. Tweets with fewer than four remaining words are excluded. After preprocessing, we construct the TF-IDF vectors using unigrams to represent the textual dimensions.

%\vspace{-1cm}
\subsection*{Interaction-Based Coordination Detection}

Although coordination is usually conceptualized as a case where two actors collude to share the same content, users may also collude by directly engaging with one another with the goal of artificially boosting their content's reach. We refer to this as \textit{interaction-based coordination}. 
Specifically, we consider if actors from one state retweet or reply to posts produced by users from another state. 

To test if interaction rates between users from a pair of states exceed what one might expect in the absence of coordination, we wish to measure the suspiciousness of engagement between users in different states.  
Our intuition is that an inactive user's engagement with a small campaign is more indicative of potential coordination than a highly active user's engagement with a large campaign.
This intuition is captured by a \textit{suspiciousness score}, based on a weighted retweet/reply network. For a given user $i$ and country $c$ (the country in the state pair to which $i$ does not belong), we define suspiciousness as $\mathrm{Score}_{i,c} = \frac{S_{i,c}}{S_i} \times \frac{S_{i,c}}{P_c}$, where ${S_{i,c}}$ is the number of retweets/replies by $i$ to users in country $c$. $S_i$ is the
out-strength (total retweets/replies) by user $i$, and $P_c$ is the number of original posts produced by users in country $c$. 
The two factors correspond to the fraction of a user's activity going to the complementary state and the fraction of content produced by the other state that the user has engaged with, respectively. In other words, suspiciousness measures the tendency of a user to engage with content from another state after taking into account the activity level of that user as well as the number of opportunities (original posts) that the other state offers for engagement.

%If for a given pair of states the test's p-value is below the 5\% significance level, we conclude that this state pair as engaging in (\textit{interaction-based coordination}).
%*NODE SUSPICIOUSNESS*

\subsection*{Statistical Test for Inter-state Coordination}

The primary exercise of this analysis is to use the interactions and similarities of control users as a baseline for determining what constitutes suspicious behavior for a given state pair. Our approach to using the control user data as a baseline is to calculate the distributions of the similarity/suspiciousness metrics for our IO and control users separately, and then apply the Mann-Whitney $U$ test \cite{mannWhitney} to determine if the IO distribution is statistically larger than the control distribution. 
%We utilize this test to assess whether a given state pair is coordinating based on a specific behavioral trace or interaction type. 

For interaction-based coordination, node suspiciousness scores are calculated for the IO users of the two states, forming the IO distribution for this test. We apply the same procedure to the control users associated with these two states, resulting in the control distribution. If the IO users are coordinating across these two states using this type of interaction, the IO distribution of node suspiciousness should be larger than the control distribution, which would be reflected by a low $p$-value. 

When applying this test to similarity-based networks, rather than focusing on nodes, the distributions are defined by the edge weights of connections between IO users and between control users. Similar to the interaction network analysis, we focus exclusively on edges connecting users from different states. 

The hypotheses for the test are as follows.
\begin{itemize}
    \item \textit{Null Hypothesis} ($H_0$): There is no significant difference between the distributions observed in coordinated users and the control group, i.e., any differences are attributable to random variation, indicating no evidence of coordination. 
    \item \textit{Alternative Hypothesis} ($H_1$): The distributions observed in coordinated users are significantly greater than those of the control group, suggesting evidence of coordination.
\end{itemize}
We evaluate each state pair according to all the similarity and interaction traces listed above, resulting in 197 experiments based on the traces observed in the data. 
% We apply a Bonferroni correction to the significance level of 0.05, resulting in a significance threshold of 0.0002 for all subsequent tests.

\section*{Evaluation}

Wang et al.~\cite{wang2023evidence} based their analysis on networks of state-state interactions aggregated across IO users. 
Fig.~\ref{fig:graphs} visualizes such aggregate networks. While Wang et al. combined retweets, quotes, replies, and mentions, we consider separate directed graphs for retweet and reply interaction networks. An edge weight represents the number of interactions among pairs of users in the two corresponding states, normalized by the source state's activity volume (aggregate out-strength). 
We build analogous networks from aggregate similarity indicators (not considered by Wang et al.). An edge weight is based on the indicators (e.g., hashtags) shared by users in the two corresponding states. To account for the activity volumes of both states, we normalize using the Jaccard coefficient, calculated as the intersection of behavioral traces divided by their union. 
An inspection of these aggregate networks reveals that certain states exhibit relatively higher edge weights across different indicators. For retweets and replies, Egypt-UAE and Qatar appear to have some form of relationship, and so do Iran and Cuba. In the similarity networks, Russia, China, and Iran seem to coordinate, among others.

To evaluate whether the patterns observed at the aggregate state-state level are truly suspicious, we apply the Mann-Whitney $U$ test 
%a null model approach using the Mann-Whitney U test at the disaggregated user level for each pair of states, as 
described in the \textit{Methods} section. 
For each interaction and similarity network, our analysis focuses on potentially coordinated state pairs, where the $U$ variable is higher for the IO network than the control network. These pairs are listed in Table~\ref{table:p_values}. 
We find that no state-state edge, in any of our networks, is statistically significant, even after adjusting the significance threshold ($p=0.05$) with a Bonferroni correction. 
For example, applying the test to the edges between Egypt-UAE and Qatar in the retweet and reply networks, which appeared suspicious at the aggregate level, yields $p=0.49$ and 0.44, respectively. These high values indicate that we cannot rule out the null hypothesis that the observed patterns occurred due to reasons other than coordination, as they are consistent with user-level observations in the control data. 
In other words, we find no statistical evidence of inter-state coordination. 
%The patterns observed at the aggregate level do not hold up under user-level analysis when incorporating the corresponding control data.

\begin{table}
   \centering
       \captionof{table}{Significance level of Mann-Whitney $U$ tests for each potentially coordinated pair of countries.}
   \begin{tabular}{|l|c|c|}
   \hline
   \textbf{Country Pair} & \textbf{\textit{p}-value}\\ \hline\hline
   %retweets
   %UAE - Egypt \& UAE & 0.008224\\ 
   \textbf{Retweets} &\\
   Qatar - Egypt \& UAE & 0.49\\
   Qatar - Bangladesh & 0.61\\
   Qatar - Catalonia & 0.61\\
   Thailand - Catalonia & 0.61\\ 
   %replies
   %Egypt \& UAE - UAE & 1.608672e-25\\
   \hline
   \textbf{Replies} & \\
   Egypt \& UAE - Qatar & 0.44\\
   Spain - Qatar & 0.56\\
   Ghana \& Nigeria - Qatar & 0.56\\
   Catalonia - Bangladesh & 0.57\\
   \hline
   \textbf{Co-Retweet} & \\
   Armenia - Russia & 0.39\\
   \hline
   \end{tabular}
    \label{table:p_values}
\end{table}

\section*{Conclusions}

Our analysis builds on prior evaluations of potential inter-state coordination by incorporating control data and state-of-the-art computational models to assess whether observed similarities or interactions between states are indicative of strategic coordination. This approach enables us to filter out spurious similarities or interactions that, while present, do not constitute evidence of coordination. Our findings indicate that the observed relationships between all state pairs analyzed are statistically insufficient to be considered signals of coordination, highlighting the need for robust methodologies and control datasets in detecting online coordination and encouraging caution when interpreting interaction patterns as signals of orchestrated campaigns.

We must acknowledge some limitations of the present analysis. The control data used in our study was only available for a subset of documented information operations, limiting our ability to generalize findings to states not represented in the control. Additionally, while the control data aims to capture relevant conversations by sampling users based on hashtag use, this heuristic could sometimes result in conversations that are either overly broad or overly narrow compared to the information operations.

\section*{Acknowledgements}
This work was supported in part by DARPA (\#HR001121C0169).

\bibliographystyle{ACM-Reference-Format}
\balance
\bibliography{references}

\end{document}